# Towards a more efficient use of process and product traceability data for continuous improvement of industrial performances

Thierno M.L. DIALLO[1], Sébastien HENRY[1], Yacine OUZROUT[2]

[1] DISP laboratory, University of Lyon, University Lyon 1, France
{Thierno.Diallo,Sebastien.Henry}@univ-lyon1.fr
[2] DISP laboratory, University of Lyon, University Lyon 2, France
Yacine.Ouzrout@univ-lyon2.fr

**Abstract.** Nowadays all industrial sectors are increasingly faced with the explosion in the amount of data. Therefore, it raises the question of the efficient use of this large amount of data. In this research work, we are concerned with process and product traceability data. In some sectors (e.g. pharmaceutical and agro-food), the collection and storage of these data are required. Beyond this constraint (regulatory and / or contractual), we are interested in the use of these data for continuous improvements of industrial performances. Two research axes were identified: product recall and responsiveness towards production hazards. For the first axis, a procedure for product recall exploiting traceability data will be propose. The development of detection and prognosis functions combining process and product data is envisaged for the second axis.

## 1    Introduction

The work presented here is part of a research project involving industrial firms and research laboratories. The purpose of this project is the setting up and management of a unitary traceability system throughout the user sector. This traceability carries on both process and products parameters and concern the entire life cycle of the product. The traceability system will ensure a serialized unique identification at the item level. The fields of application referred are characterized by complex and difficult to model process, a large number of parameters and a high variability. Disruptions in product flows are also observed in several places within supply chain (FIFO broken).The amount of data gathered is therefore very important. Regarding the work package presented in this research proposal, the goal is to use the data collected for the continuous improvement of industrial performances. These improvements will focus on both production and supply chain. The particularity of this project is twofold. First, in the process industries, generally the performed traceability is a material traceability relating to production lots. In our case, we are targeting a unitary traceability enabling a serialized unique identification at the product item level. The aim is to connect to each product the exact parameters values of its production processes and events of its life cycle. The second particularity lies in the use of traceability data. The use that is made of such data by businesses is mainly to protect themselves in case of incidents or to respond to a request from the authorities or customers. As this type of traceability allows a more detailed view of the process and the product life cycle, we wish to exploit its potential for continuous improvement. It will be, from the identifier of a process or a product, to compare observed traceability data with desired master data. When discrepancies are found, propose corrective actions on the controlled process or on the products flow (see fig.1).

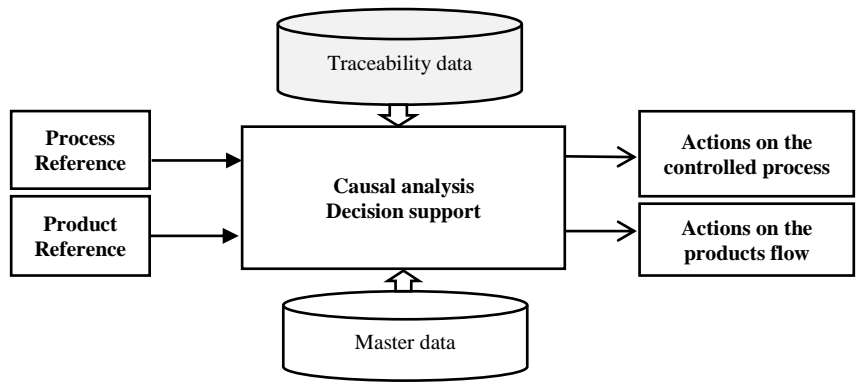

**Fig. 1**: The potential use of unitary traceability data

The collection and management of such data are important technical and organizational challenges. In this work package we focus on how to exploit the richness and the fineness of the data collected by the traceability system. The coupling between product and processes data will allow better analysis and understanding of the existing relationships between process parameters and product quality.

This research proposal is organized as follow. In Section 2, we make the research problem statement. The literature review and the research gaps are provided in Section 3. The research questions addressed and research methodology are presented in Section 4 and Section 5 respectively. Section 6 enumerates the expected outputs of this research project. In Section 7, we display the timetable of the overall research project.

## 2 Problem statement

Generally, the main motivation for setting up a traceability system is related to regulatory and contractual requirements. Large amounts of data related to process and product parameters are thus collected and stored. Companies have recourse to these data in case of incidents or requests from government or its partners. Despite the potential of these data, they are rarely used for the improvement of industrial processes. This research project was initiated from an industrial need concerning the use of traceability data in a continuous improvement process. As the potential use of this data is huge, we have sought to delimit the scope of our research. To do this, we have established with industrialists a functional specifications related to our work package. The aim was to clearly define the industrial needs and to prioritize them. From the main industrial issue which was to minimize direct and indirect costs of non-quality, we have deduced seven use cases (see Fig.2). Four was located within the value chain and three outside the value chain. Within the value chain, the goal is to minimize the number of products sold with a defect. This can be achieved by improving the production process and the enhancement of quality control. When the products are outside the value chain and that a noncompliance requiring a recall is found, we wish to find the recall procedure minimizing the direct and indirect financial impacts.

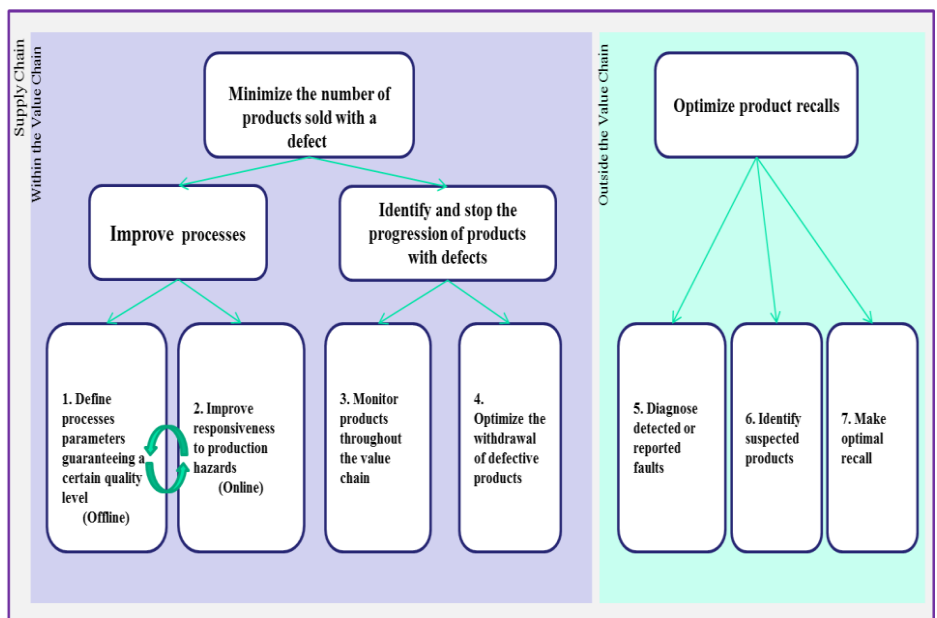

**Fig.2**: Expected use cases of our work package

From these use cases, three research areas were identified: traceability, product recall procedure and FDI (Fault Detection and Identification) functions. In order to determine the state of the art and identify research gaps, we have conducted a preliminary literature review.

## 3 Preliminary literature review and research gaps

In this preliminary research work, we have critically reviewed the three research areas in relation with our research project (traceability, product recall procedure and FDI functions).

### 3.1 Traceability

Traceability is an interdisciplinary research field which spans fundamental sciences as well as the management sciences (economy, supply chain, production, quality, marketing, etc.) (Karlsen et al., 2013). It originated in the mid-1980s in logistics to control the flow of goods within a supply chain (Schiffers, 2011). It applies nowadays at almost all industrial sectors (pharmaceutical, agribusiness, automotive industry, etc.) and for varied purposes (compliance with regulatory constraints, chain communication, marketing, etc.). The concept of traceability can have different meaning depending on the scope and the intended use. In the literature review carry out by Karlsen, K.M., et al (Karlsen et al., 2013), the authors conclude that there is no common understanding of the definitions and principles of traceability. Olsen, P. and M. Borit (Olsen and Borit, 2013) make an overview of relevant traceability definitions, outlining similarities and differences. The authors found that the descriptions and properties of a traceability system provided by the articles studied are virtually identical. The main proprieties identified was:
- Ingredients and raw materials must be grouped into units with similar properties (notion of "Traceable Resource Units")
- Identifiers/keys must be assigned to these units
- Product and process properties must be recorded and either directly or indirectly linked to these identifiers
- A mechanism must exist to get access to these properties

The authors use these traceability system properties as a benchmark to evaluate the traceability definitions and found that a correct definition of traceability does not currently exist. The authors then propose a generic definition of traceability.

**Definition (Traceability):** *The ability to access any or all information relating to that which is under consideration, throughout its entire life cycle, by means of recorded identifications* (Olsen and Borit, 2013).

We distinguish two levels of traceability: internal traceability (within the considered chain value) and chain traceability (outside the chain value).

To access an information, it should be beforehand defined and then recorded. It is impossible to get information about a product property for example if it's not clearly identified and his value stored. However, there is no general rule on what data to collect. The nature and organization of data to be collected will therefore be one of the issues to be addressed.

The different food scandals have fact that traceability is more and more required by the regulation. Beyond this regulatory aspect, applications of traceability in food industries include production optimization, competitive strategy and increase of company coordination in supply chain (Ruiz-Garcia et al., 2010, Moe, 1998, Galvão et al., 2010). Our aim is to exploit this second possibilities offered by traceability. In particular, we want to optimize product recall and increase responsiveness through traceability data.

**Research gaps:** Current scientific work is mainly oriented towards the external traceability. This can be explained by the origin of traceability which was first applied to the management of the supply chain. At this external level, information collected and shared focus primarily on events and identifiers of objects and places. This allows to track a product throughout the supply chain. Traceability is also performed internally by different services, but with different methods and objectives. For example, material traceability allows to know the lots of raw material used in production and machines historical is used for maintenance. But these data are usually managed by different systems and reconciliation are not made between them. The connections among all these data in a continuous improvement process will be studied in this research.

### 3.2 Product recall

Product recall is *any measure aimed at achieving the return of a dangerous product that has already been supplied or made available to consumers by the producer or distributor*. It should be made the difference between product recall and product withdrawal. The latter is defined *as any measure aimed at preventing the distribution, display and offer of a product dangerous to the consumer* (EU, 2001). According to the degree of dangerousness of the product, three classes of recall are distinguished (Kumar and Budin, 2006) :

Class 1: This is the more stringent class. It is advocated when the use or exposure to the product can cause serious and lasting health problems or death.
Class 2: The product may cause temporary health problems but can lead in the long-term to serious problems.
Class 3: With the lowest severity, it concerns cases where there is no health risk.

The strategy and the impact of the recall obviously vary according to the class concerned. Government agencies define mandatory to follow in case of a recall procedures in the UE (services, 2014), the USA (Commission, 2014, Administration, 2014), and Australia (Branch and Commission). Some industrial standards provide guidance on how to manage a product recall process (See the one proposed by GS1 (GS1, 2012)). In the annual reports produced by the agencies in charge of consumer safety, we are seeing a steady increase in the number of dangerous products reported (see for example (Commission, 2013b), (Commission, 2013a) and (Kramer et al., 2005).

Despite the challenges that may have a recall problem, the scientific literature in this area is not very rich (Magno, 2012). Existing studies often focus on the impacts (financial, brand image ...) that these recalls may have and the management of this type of crisis(Cheah et al., 2007) (Magno, 2012). A few studies have been published on optimizing the recall process. In (Kumar, 2014), Failure Mode Effects and Criticality Analysis (FMECA) and fault tree studies are used to determine the causes of noncompliance at the origin of the recall and assess the reliability of the recall supply chain. Kumar, S. and E.M. Budin(Kumar and Budin, 2006) propose an approach based on the HACCP method to prevent recalls or better manage the crisis caused by a recall. Piramuthu, S. et al. (Piramuthu et al., 2013) propose a probabilistic model of the contaminated place to estimate the cost for each actor in a supply chain upon recall. They consider a supply chain consisting of three levels: producer, distributor and retailer. Conze, D.B. and C. Kruger(Conze and Kruger, 2013) define the recall strategy to adopt based on a quantitative probabilistic risk analysis. The work published in (Chen and Schweickert, 2004) intend to determine the conditional probability of recall of a product knowing that the products just before or after are recalled. To reduce the size of the recalled lots, other authors have proposed to reduce the dispersion of the final product by reducing the size and the mixing of production lots using either linear programming (Dupuy et al., 2005)or neural networks and genetic algorithms (Tamayo et al., 2009).

In our view, in order to optimize the recall, one must succeed in determining the origin of the noncompliance as soon as possible. Reducing the dispersion proposed in (Dupuy et al., 2005) and (Tamayo et al., 2009) is not always feasible because it is often induced by the manufacturing process and the supply and distribution policy.

**Research gaps:** When a noncompliance is detected and a recall is required, conventional recall procedure the strategy usually adopted consists of recalling all the lot to which the detected nonconformity belongs. The recall is therefore done without really knowing the root causes of the noncompliance. However this search for root causes before performing the recall allows to limit the recall to only faulty items and to correct the sources of noncompliance. This search for causes must be done without loss of time due to the need to react quickly to this type of crisis. It is therefore necessary to be prepared in advance. The few studies on finding the root cause in the case of a product recall procedure that we have consulted employ deterministic (FMECA, HACCP, fault tree, etc. ) methods for modeling causality. However, the use of these methods in industrial contexts characterized by complex processes and a large number of parameters is not always possible. In this research work, we will investigate other alternatives for modeling causal relationships between process parameters and non-compliances.

### 3.3 Fault Detection and Isolation

Fault Detection and Isolation (FDI) consists of monitoring a system, identifying when a fault has occurred, and pinpointing the type of fault and its location (Wikipedia, 2014). The reactivity will depend largely on the ability to detect and to diagnose failure or degradation. The diagnosis may also include the prognosis. The detection allows to signal the occurrence of a fault in a given system (Fonctionnement, 2000, Isermann and Ballé, 1997). It consists in observing the parameters or characteristics of the studied system to ensure that they have acceptable values. The diagnose function allows to isolate and identify the detected fault (Isermann, 1984, Isermann and Ballé, 1997) (Gertler, 1988). One of the conditions to diagnose a fault is the knowledge of its symptoms. It is therefore essential to establish a relationship between observations and faults (Isermann, 1997, Venkatasubramanian et al., 2003b).

There are generally two FDI approaches (Frank, 1990, Gertler, 1988): model-based FDI and knowledge-based FDI. The first approach use process and signal models and the second one is based on analytical and heuristic information (Isermann, 1997). It is difficult to obtain a reliable model for industrial systems because the processes are often complex and non-linear (Venkatasubramanian et al., 2003a). The second approach allows to determine the operating mode of the system based on historized observations on the system and flows (Dubuisson, 1990, Jain et al., 2000). The application of this approach still encounters many difficulties such as the determination of the minimal set of parameters modeling the actual functioning of the system and the generation to new situations. Most often, the FDI developed approaches use process data. In this research work, we will combine process and product parameters data to make FDI functions more accurate.

**Research gaps:** As mentioned previously the success of the process of diagnosing a fault depends on the ability to recognize its symptom. But in some cases, several defaults may have the same symptom. Thus arises the question on how to isolate a particular default. The use of product parameters data in addition to those of the machine parameters for diagnosis of multiple faults will be addressed in this work.

## 4  Research questions

Based on the problem statement and the literature review, we intend to address the following research questions:
- What data to collect and how to organize them?
- How traceability can contribute in to product recall optimization and responsiveness rising?
- How and to what extent the combination of process and product traceability data can contribute to the detection and diagnosis of multiple and unobservable defects?

## 5  Research methodology

To carrying out this research project, we will achieve successively the following tasks.

### 5.1  State of the art

This first step is the pursuit of the literature review to identify the existing theoretical contributions to the research areas of interest. The existing literature will be analysed and the research gaps regarding the needed use cases will be determine. Some databases (ProQuest Dialog, ScienceDirect, Web of Knowledge, and Google Scholar) and keywords have already been chosen. Some research communities and labs working on research areas of interest have been identified.

## 5.2 Data collection

This task can be performed simultaneously with the previous one. It allows to collect required data and to interview future users. The collected data include process and product historical data. Sufficient knowledge of industrial processes shall also be acquired.

## 5.3 Formulation of a hypothesis solution

From the research gaps, some hypothesis will be done on how the unitary product and process traceability data can contribute to meet industrial needs. These assumptions will be confronted to the industrial context and criticism of other researchers.

## 5.4 Research problems solving

At this stage, we will propose answers to research questions. This resolution step will involve the following tasks:
- Modelling the industrial processes: models are required for FDI functions
- Developing the causal analysis framework: this framework will allow establish cause-effect relationships between process/product parameters and defects.
- Developing algorithms for failure detection, prognosis and recall process: These algorithms will describe procedures to be followed to achieve the different use cases.
- Simulation: This is to validate the proposed approaches, algorithms and functions.
- Pilot implementation: The validated solutions will be implemented in industrial context.

# 6 Expected results and contributions to theory and practice

The expected results of this research work are both theoretical and practical. They include:
- State of the art of FDI functions
- Data model for unitary traceability
- Proposal of a product recall procedure
- Causal analysis algorithms for detection, diagnosis and prognosis

## 7 Research timetable

We have developed a timetable in order to define stages of the overall research project and their corresponding ongoing goal. It also allows to follow the progress of the project and to react quickly in case of delay.

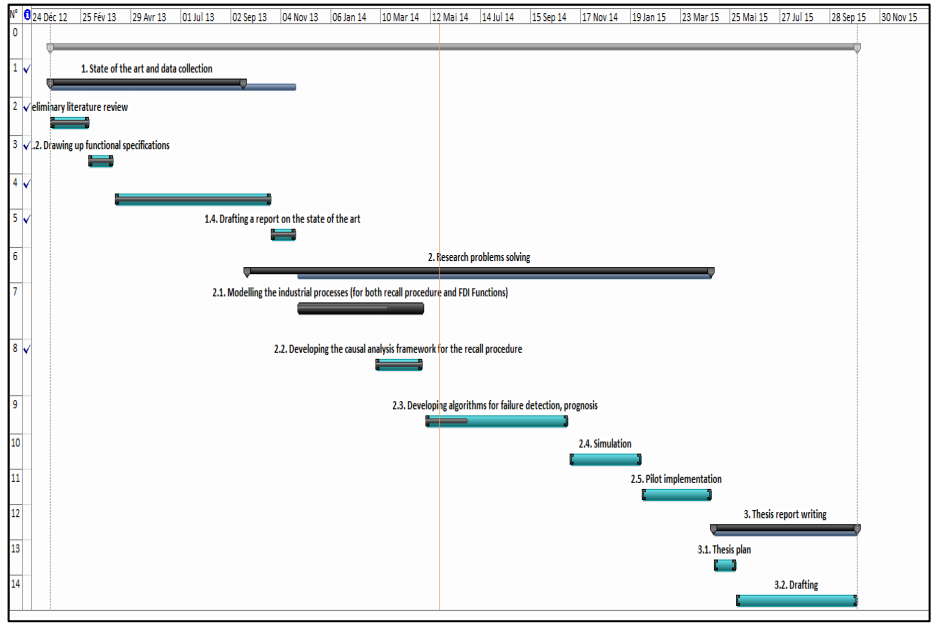

**Fig.3**: Our research project planning